\documentclass{article}

\usepackage{amssymb,amsfonts,amsmath}
\usepackage{cite,enumerate,float,indentfirst}
\usepackage{color}

\def\be{\begin{eqnarray}}
\def\ee{\end{eqnarray}}
\def\nn{\nonumber}

\hoffset= - 1.0in         

\begin{document}

\hfill ITEP/TH-01/18

\hfill IITP/TH-01/18

\bigskip

\centerline{\Large{Knot polynomials for twist satellites
}}

\bigskip

\bigskip

\centerline{\bf  A.Morozov }

\bigskip

{\footnotesize
\centerline{{\it
ITEP, Moscow 117218, Russia}}

\centerline{{\it
Institute for Information Transmission Problems, Moscow 127994, Russia
}}

\centerline{{\it
National Research Nuclear University MEPhI, Moscow 115409, Russia
}}
}

\bigskip

\bigskip

\centerline{ABSTRACT}

\bigskip

{\footnotesize
We begin the systematic study of knot polynomials for the twist satellites of
a knot, when its strand is substituted by a 2-strand twist knot.
This is a generalization of cabling (torus satellites),
when the substitute of the strand was a torus knot.
We describe a general decomposition of satellite's colored HOMFLY in those
of the original knot, where contributing are
adjoint  and other representations from the ``$E_8$-sector'',
what makes the story closely related to Vogel's universality.
We also point out a problem with lifting the decomposition rule to
the level of superpolynomials -- it looks like such rule, if any, should be different
for positive and negative twistings.
}

\bigskip

\bigskip

\tableofcontents

\section{Introduction}

Knot polynomials \cite{knotpols}
are the only observables in the Chern-Simons theory \cite{CS}.
They are at once the simplest among the Wilson loop averages in gauge theories
and among explicit realizations of dualities (modular transformations)
in conformal theories.
They are essentially non-perturbative and reveal the relevance of
topological-field-theory description of non-perturbative physics.
They can be explicitly calculated -- and thus provide a set of exactly solvable
examples for all above-mentioned topics, which are undoubtedly the central ones
of today's theoretical physics.

Still, the theory of knot polynomials is making just its first steps,
and explicit calculations are still available only for very restricted
classes of knots.
Any extension of these classes is valuable and can help to identify the
general properties of physical observables and build their alternative
descriptions, which do not refer to gauge theory formalism and can be used
in confinement phases.
In this letter we suggest one of such awaited extensions:
from any previously-studied knot to its $k$-twisted satellite,
which is usually a knot of a far more complicated structure.
Still, as we demonstrate, its knot polynomials can be directly reduced
to those of the original knot, i.e. we can actually build an infinite sequence
$$
{\rm knot}\ \  \longrightarrow \ \ {\rm its\ satellites} \ \
\longrightarrow \ \  {\rm satellites\ of\ these\ satellites} \ \
\longrightarrow \ \ldots
$$
leading to knots of arbitrary high complexity.
This letter is just the first step, but the idea should be clear --
as well as the many challenges and problems it can raise and resolve.

\section{Whitehead doubles}

The $k$-twist satellite, also known as Whitehead double, 
 $S_k({\cal K})$ of ${\cal K}$ is defined \cite{sat}
as the image of the twist knot ${\rm Twist}_k$,
naturally embedded into a $3d$ solid torus,
which is then identified with the tubular neighborhood of the knot ${\cal K}$:

\begin{picture}(300,310)(-80,-180)

\put(15,100){

\qbezier(-45,-58)(-40,-58)(-35,-55)\qbezier(-25,-51)(-30,-51)(-35,-55)
\qbezier(-49,-62)(-40,-62)(-35,-65)\qbezier(-25,-69)(-30,-69)(-35,-65)

\qbezier(45,-58)(40,-58)(35,-55)\qbezier(25,-51)(30,-51)(35,-55)
\qbezier(49,-62)(40,-62)(35,-65)\qbezier(25,-69)(30,-69)(35,-65)

\put(-35,-55){\vector(1,1){2}}
\put(-35,-65){\vector(-1,1){2}}
\put(5,-51){\vector(1,0){2}}
\put(3,-69){\vector(-1,0){2}}
\put(35,-55){\vector(1,-1){2}}
\put(35,-65){\vector(-1,-1){2}}

\qbezier(25,-51)(24,-51)(22,-52)\qbezier(25,-69)(8,-69)(17,-57)
\qbezier(0,-69)(11,-69)(13,-68)\qbezier(0,-51)(30,-51)(18,-65)

\put(-25,-75){\line(0,1){30}}
\put(0,-75){\line(0,1){30}}
\put(-25,-75){\line(1,0){25}}
\put(-25,-45){\line(1,0){25}}

\put(-21,-65){\mbox{${\cal R}^{2k}$}}

\qbezier(-45,-58)(-70,-58)(-70,-37) \qbezier(-45,-16)(-70,-16)(-70,-37)
\qbezier(-49,-62)(-74,-62)(-74,-37) \qbezier(-49,-12)(-74,-12)(-74,-37)
\qbezier(45,-58)(70,-58)(70,-37) \qbezier(45,-16)(70,-16)(70,-37)
\qbezier(49,-62)(74,-62)(74,-37) \qbezier(49,-12)(74,-12)(74,-37)
\put(-45,-16){\line(1,0){90}}
\put(-49,-12){\line(1,0){98}}

\qbezier(-48,-31)(0,-50)(48,-31)
\qbezier(-40,-33)(0,-15)(40,-33)
\qbezier(-90,-51)(0,-121)(90,-51)
\qbezier(-90,-25)(0,45)(90,-25)
\qbezier(-90,-51)(-110,-38)(-90,-25)
\qbezier(90,-51)(110,-38)(90,-25)

\put(-15,-5){\mbox{${\rm Twist}_k$}}

}

\put(10,-60){

\qbezier(-48,5)(-30,-65)(50,0)
\qbezier(20,42)(100,47)(50,0)
\qbezier(0,40)(-20,38)(-60,3)
\qbezier(-49,20)(-45,90)(52,10)

\qbezier(60,3)(100,-35)(50,-53) \qbezier(-60,3)(-100,-35)(-50,-53)
\qbezier(-50,-53)(0,-70)(50,-53)

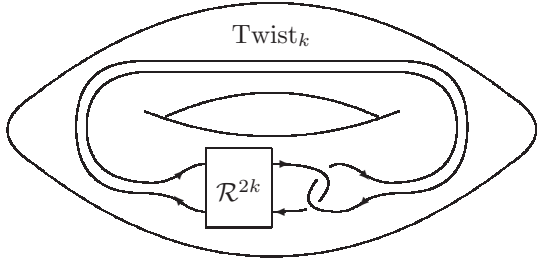
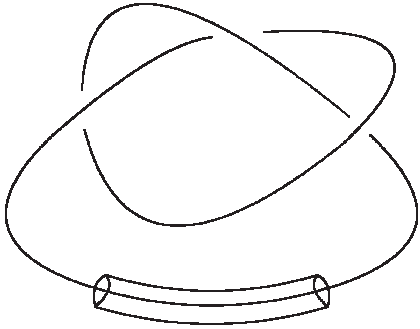
\put(-30,-90){\mbox{${\cal K}=3_1=$ trefoil}}
\put(-80,-102){\mbox{with a fragment of its tubular neighborhood}}

\qbezier(-40,-50)(0,-62)(40,-50)
\qbezier(-44,-62)(0,-75)(44,-62)
\qbezier(-40,-50)(-46,-53)(-44,-62)
\qbezier(-40,-50)(-36,-55)(-44,-62)
\qbezier(40,-50)(46,-53)(44,-62)
\qbezier(40,-50)(36,-55)(44,-62)
}

\put(250,0){

\qbezier(-44,12)(-30,-65)(50,5)     \qbezier(-48,9)(-30,-69)(50,0)
\qbezier(20,42)(110,51)(50,0)      \qbezier(24,39)(100,48)(50,5)
\qbezier(0,40)(-20,38)(-60,12)      \qbezier(6,37)(-20,34)(-60,8)
\qbezier(-49,22)(-48,90)(52,16)    \qbezier(-45,24)(-45,84)(50,12)
\qbezier(-60,8)(-132,-35)(-55,-58) \qbezier(-60,12)(-140,-35)(-55,-62)
\qbezier(61,5)(125,-35)(55,-58)    \qbezier(64,8)(132,-35)(55,-62)

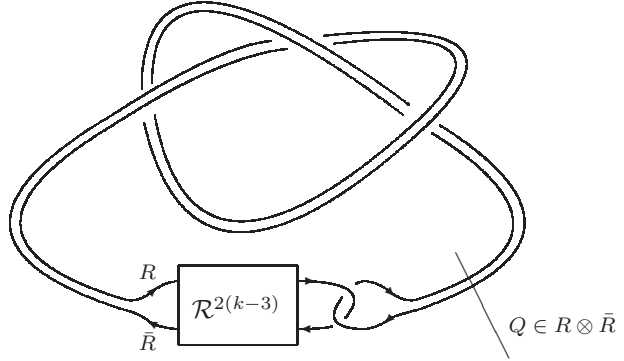
\put(-65,-105){\mbox{$S_k(3_1)= k-$twist satellite of the trefoil}}
\put(-10,-119){\mbox{with $w^{3_1'} = -3$}}

\put(10,0){

\qbezier(-65,-58)(-60,-59)(-55,-55)\qbezier(-45,-51)(-50,-51)(-55,-55)
\qbezier(-65,-62)(-60,-63)(-55,-67)\qbezier(-45,-69)(-50,-69)(-55,-67)

\qbezier(45,-58)(40,-59)(35,-55)\qbezier(25,-51)(30,-51)(35,-55)
\qbezier(45,-62)(40,-63)(35,-65)\qbezier(25,-69)(30,-69)(35,-65)

\put(-55,-55){\vector(1,1){2}}
\put(-55,-67){\vector(-2,1){2}}
\put(5,-51){\vector(1,0){2}}
\put(3,-69){\vector(-1,0){2}}
\put(35,-55){\vector(1,-1){2}}
\put(35,-65){\vector(-1,-1){2}}

\put(-60,-50){\mbox{\footnotesize $R$}}
\put(-60,-77){\mbox{\footnotesize $\bar R$}}
\put(80,-70){\mbox{\footnotesize $Q\in R\otimes\bar R$}}
\put(60,-40){\line(1,-2){20}}

\qbezier(25,-51)(24,-51)(22,-52)\qbezier(25,-69)(8,-69)(17,-57)
\qbezier(0,-69)(11,-69)(13,-68)\qbezier(0,-51)(30,-51)(18,-65)

\put(-45,-75){\line(0,1){30}}
\put(0,-75){\line(0,1){30}}
\put(-45,-75){\line(1,0){45}}
\put(-45,-45){\line(1,0){45}}

\put(-40,-65){\mbox{${\cal R}^{2(k-3)}$}}


}
}

\end{picture}

\noindent
${\rm Twist}_k$ and ${\cal K}$ are often called respectively 
the {\it pattern} and the {\it companion}
of the satellite $S_k({\cal K})$. 
Likewise one can define satellites of other types, 
e.g. with patterns which are torus knots
(satellites in this case would be just the ordinary {\it cablings} \cite{cabling}
of ${\cal K}$),
but in this letter we concentrate on the $k$-twist ones.

\section{HOMFLY polynomials for satellites}

According to \cite{MMM2}, reduced HOMFLY polynomial
in representation $Q$ for a knot ${\cal K}$,
described as a closure of an $m$-strand braid
is given by
\be
{\cal H}^{\cal K}_Q = \sum_{Y\in Q^{\otimes m}}
D_Y\cdot {\rm Tr}_{{\rm mult}_{_Y}} {\cal R}_Y^{\cal K}
\ee
where ${\cal R}_Y$ is a convolution of ${\cal R}$-matrices in
Tanaka-Krein representation, which are square matrices of the size
${\rm mult}_{_Y}\times {\rm mult}_{_Y}$ 
and ${\rm mult}_{_Y}$ is the multiplicity
of representation $Y$ in the decomposition of the product
\be
Q^{\otimes m} = \oplus_{_Y} {\rm mult}_{_Y}\cdot Y
\ee

Now, the satellite $S_k({\cal K})$ is obtained
by substituting the knot by a 2-wire antiparallel cable,
carrying representation $R\otimes \bar R$, and by changing
one of the $m$ traces for a cut of the twist knot ${\rm Twist}_{k+w^{\cal K'}}$.
Here $w^{\cal K'}$ is the writhe number of ${\cal K'}$,
which actually depends on the $2d$ knot diagram, not just on the knot --
we put prime over ${\cal K}$ to remind about this diagram dependence.
In what follows we often write just $w$ instead of $w^{\cal K'}$
to simplify the formulas.
Strictly speaking the satellite itself also depends on the
knot diagram, but   this dependence can be compensated by the shift of $k$,
see (\ref{twotrefs}) below for a simple example --
and we omit prime in $S_k({\cal K}')$.

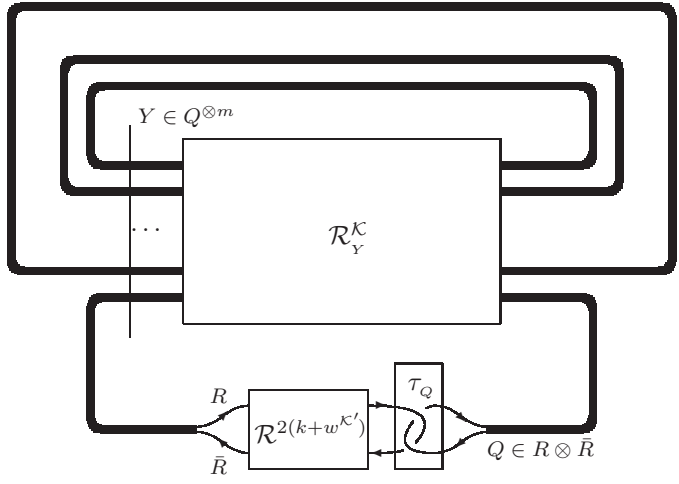
\begin{picture}(300,210)(-220,-90)

\put(-70,50){\line(1,0){120}}
\put(-70,-20){\line(1,0){120}}
\put(-70,50){\line(0,-1){70}}
\put(50,-20){\line(0,1){70}}
\put(-15,10){\mbox{${\cal R}^{\cal K}_{_Y}$}}
\put(-90,-25){\line(0,1){80}}
\put(-87,56){\mbox{\footnotesize $Y \in  Q^{\otimes m}$}}

\qbezier(-65,-59)(-60,-59)(-55,-55)\qbezier(-45,-51)(-50,-51)(-55,-55)
\qbezier(-65,-61)(-60,-61)(-55,-65)\qbezier(-45,-69)(-50,-69)(-55,-65)

\qbezier(45,-59)(40,-59)(35,-55)\qbezier(25,-51)(30,-51)(35,-55)
\qbezier(45,-61)(40,-61)(35,-65)\qbezier(25,-69)(30,-69)(35,-65)

\put(-55,-55){\vector(1,1){2}}
\put(-55,-65){\vector(-1,1){2}}
\put(5,-51){\vector(1,0){2}}
\put(3,-69){\vector(-1,0){2}}
\put(35,-55){\vector(1,-1){2}}
\put(35,-65){\vector(-1,-1){2}}

\put(-60,-50){\mbox{\footnotesize $R$}}
\put(-60,-77){\mbox{\footnotesize $\bar R$}}
\put(45,-70){\mbox{\footnotesize $Q\in R\otimes\bar R$}}

\qbezier(25,-51)(24,-51)(22,-52)\qbezier(25,-69)(8,-69)(17,-57)
\qbezier(0,-69)(11,-69)(13,-68)\qbezier(0,-51)(30,-51)(18,-65)

\put(-45,-75){\line(0,1){30}}
\put(0,-75){\line(0,1){30}}
\put(-45,-75){\line(1,0){45}}
\put(-45,-45){\line(1,0){45}}

\put(-43,-65){\mbox{${\cal R}^{2(k+w^{\cal K'}\!)}$}}

\put(10,-75){\line(0,1){40}}
\put(28,-75){\line(0,1){40}}
\put(10,-75){\line(1,0){18}}
\put(10,-35){\line(1,0){18}}
\put(15,-45){\mbox{$\tau_{_Q}$}}

\linethickness{1mm}
\put(-100,40){\line(1,0){30}}  \put(50,40){\line(1,0){30}}
\put(-110,30){\line(1,0){40}}  \put(50,30){\line(1,0){40}}
\put(-90,15){\mbox{$\ldots$}}
\put(-130,0){\line(1,0){60}}   \put(50,0){\line(1,0){60}}
\put(-100,-10){\line(1,0){30}}   \put(50,-10){\line(1,0){30}}

\put(-100,70){\line(1,0){180}}
\put(-110,80){\line(1,0){200}}
\put(-130,100){\line(1,0){240}}

\put(-100,-60){\line(1,0){35}}  \put(45,-60){\line(1,0){35}}

\qbezier(-100,40)(-105,40)(-105,45)  \qbezier(-100,70)(-105,70)(-105,65)
\put(-105,45){\line(0,1){20}}
\qbezier(-110,30)(-115,30)(-115,35)  \qbezier(-110,80)(-115,80)(-115,75)
\put(-115,35){\line(0,1){40}}
\qbezier(-130,0)(-135,0)(-135,5)  \qbezier(-130,100)(-135,100)(-135,95)
\put(-135,5){\line(0,1){90}}
\qbezier(-100,-10)(-105,-10)(-105,-15)  \qbezier(-100,-60)(-105,-60)(-105,-55)
\put(-105,-55){\line(0,1){40}}

\qbezier(80,40)(85,40)(85,45)  \qbezier(80,70)(85,70)(85,65)
\put(85,45){\line(0,1){20}}
\qbezier(90,30)(95,30)(95,35)  \qbezier(90,80)(95,80)(95,75)
\put(95,35){\line(0,1){40}}
\qbezier(110,0)(115,0)(115,5)  \qbezier(110,100)(115,100)(115,95)
\put(115,5){\line(0,1){90}}
\qbezier(80,-10)(85,-10)(85,-15)  \qbezier(80,-60)(85,-60)(85,-55)
\put(85,-55){\line(0,1){40}}

\end{picture}

\noindent
From this picture we read:
\be
\boxed{
H^{S_k({\cal K})}_{_R} =
\frac{1}{D_{_R}}\sum_{Q\in R\otimes \bar R}
D_{_Q}\cdot {\cal H}_{_Q}^{\cal K}\cdot \mu_{_Q}^{2(k+w^{\cal K'})} \cdot\tau_{_Q}
= \sum_{Q\in R\otimes \bar R} \alpha_{_{R,Q}}\mu_{_Q}^{2(k+w^{\cal K'}) } 
\cdot {\cal H}_{_Q}^{\cal K}
}
\label{SatH}
\ee
For ${\cal K} = {\rm unknot}$ with ${\cal H}^{\rm unknot}=1$ eq.(\ref{SatH})
reduces to the evolution formula from \cite{evo}
\be
H^{S_k({\rm unknot})}_{_R} =
H^{{\rm Twist}_k}_{_R} =
\frac{1}{D_{_R}}\sum_{Q\in R\otimes \bar R}  D_{_Q}  \cdot \mu_{_Q}^{2k} \cdot\tau_{_Q}
= \sum_{Q\in R\otimes \bar R} \!\! \alpha_{_{R,Q}}\mu_{_Q}^{2k }
\label{satunknot}
\ee

The delicate point, however, is that representation $\bar R$ and thus the emerging $Q$
explicitly depend on $N$, which parameterizes the gauge algebra $sl_N$.
In this situation the definition of $A$-dependent polynomials, like HOMFLY,
requires additional care.
Following \cite{univ1} we define them as {\it uniform} polynomials ${\cal H}$,
with the property
\be
H_{[21^{N-2}]}^{\cal K}(q,A=q^N) = {\cal H}_{\rm adjoint}(q,A=q^N)
\ee
and similarly for all other $Q\in R\otimes \bar R$.
The point is that $A$-dependence of ${\cal H}_Q$ captures both the $A$-dependence of
$H_Q(A)$ and $N$-dependence of $Q$.
In result, for example, the special and Alexander polynomials \cite{DMMSS}
\be
{\cal H}_{\rm adjoint}(q=1,A) =  H_{[1]}(q=1,A)^{2}\,,\ \ \ \ \ \ {\rm while} \ \ \ \ \ \
H_{[21^{N-2}]}(q=1,A) = H_{[1]}(q=1,A)^{N}
\ee
and
\vspace{-0.3cm}
\be
{\cal H}_{\rm adjoint}(q,A=1) =  1\,,\ \ \ \ \ \ {\rm while} \ \ \ \ \ \
H_{[21^{N-2}]}(q,A=1) = H_{[1]}(q^N,A=1)
\ee
see \cite{univ1} for further details.
Since $H_{_R}^{S_k({\cal K})}$ at the l.h.s. of (\ref{SatH}) is
well defined, appearance of the {\it uniform} HOMFLY at the r.h.s.
makes the identity a little less obvious and calls for independent checks.

\section{Fundamental HOMFLY for $S_k({\cal K})$ via the uniform adjoint for ${\cal K}$}

For the fundamental representation $R=[1]=\Box$ there are just two
$Q\in [1]\otimes\overline{[1]} = \emptyset \oplus {\rm adjoint}$.
Substituting the eigenvalues $\mu_Q$ and the matrix element $\tau_Q$
from eq.(89) of \cite{evo}, we get:
\be
\!\!\!\!\! 
\boxed{
H^{S_k({\cal K})}_{\Box} =}\ 
\overbrace{
 A^2 - \frac{A\{q\}^2}{\{A\}} - \frac{A\{Aq\}\{A/q\}}{\{A\}}\cdot A^{2(k+w )}
 }^{H^{{\rm Twist}_{k+w  }}_{_\Box}} \cdot
 {\cal H}_{adjoint}^{\cal K}
= H^{{\rm Twist}_{k+w }}_{_\Box}  
- \frac{\{Aq\}\{A/q\}}{1-A^{-2}}\cdot A^{2(k+w )}
\cdot ({\cal H}_{adjoint}^{\cal K}-1) =
\nn 
\ee
\be
\boxed{
= H^{{\rm Twist}_{k+w }}_{_\Box}
+ \Big(H^{{\rm Twist}_{k+w }}_{_\Box} - 1\Big)\Big({\cal H}_{adjoint}^{\cal K}-1\Big) 
+ {\{Aq\}\{A/q\}}{(1-A^2)}
\cdot \frac{{\cal H}_{adjoint}^{\cal K}-1}{\{A\}^2}
}
\label{satfvsadj}
\ee
where $\{x\} = x-x^{-1}$ and
the {\it uniform adjoint} \cite{univ1} HOMFLY polynomials ${\cal H}_{{\rm adjoint}}^{\cal K}$
for many simple knots are available from
\cite{univ1,univ2,knotebook}.

\section{Uniform adjoints of  twist knots}

As the simplest examples, 
the {\bf uniform adjoints}  for the trefoil  and the figure eight knots are
\be
{\cal H}_{\rm adjoint}^{3_1}
= \overbrace{
1 + \{A\}^2\cdot\Big((A^6-2A^4-A^2)}^{A^4(A^2-2)^2}
+ \ \{q\}^2 \cdot (A^8+4A^6) + \ \{q\}^4\cdot A^6\Big)
\nn\\ 
\nn \\
{\cal H}_{\rm adjoint}^{4_1}
=\underbrace{
1+\{A\}^2\cdot\Big((A^2+A^{-2})}_{(A^2-1+A^{-2})^2}
 + \ \{q\}^2\cdot (2A^2+3+2A^{-2}) + \
\{q\}^4\cdot (A^2+4+A^{-2}) + \ \{q\}^6\Big)
\label{adjH341}
\ee
Refs.\cite{univ1} and \cite{univ2} describe  evaluation of these polynomials 
respectively for arbitrary torus and arborescent \cite{arbor} knots.
In this letter  we consider just two simplest knot families:   2-strand torus 
$\ {\rm Torus}_{[2,2p+1]}$ and twist $\ {\rm Twist}_{p}$.
We denote the evolution parameters by $p$ instead of $k$, because
these knots will play the role of companions, 
while $k$ was used to parameterize the pattern.
Both families are handled by the evolution method \cite{DMMSS,evo},
which in the case of torus families leads to standard the Rosso-Jones formula
\cite{RJ,DMMSS}.
Evolution in parameter $p$ in both cases --
along the 2-strand twist and  torus families -- are described by the sums
over seven representations from the square of adjoint 
\be
{\rm adjoint}^{\otimes 2} = [21^{N-2}]\otimes [21^{N-2}]
= [1]_+ \oplus [21^{N-2}]_+ \oplus [221^{N-4}]_+ \oplus [42^{N-2}]_+ \oplus 
[21^{N-2}]_-\oplus [31^{N-3}]_-\oplus [332^{N-3}]_-
\nn
\ee
with the  eigenvalues
\vspace{-0.3cm}
\be
\Big\{\mu_Q \, \Big|\, {Q\in {\rm adjoint}^{\otimes 2}}\Big\} 
= (1,A,q^{-2}A^2,q^{2}A^2,-A,-A^2,-A^2)
\label{evs}
\ee
Pluses and minuses mark representations from symmetric and antisymmetric squares
respectively -- this affects the signs of the eigenvalues.
Since adjoint is self-conjugate there is no difference between
${\rm adjoint} {\otimes }\overline{\rm adjoint}$ and
${\rm adjoint}^{\otimes 2}$,
associated with the antiparallel (twist) and  parallel (torus) strands.
The difference between the two evolutions is that in the torus case
the weights in the sums are just dimensions, thus the contributions of two 
adjoints differ only by signs and cancel each other.
Also in the torus case eigenvalues are inverted $\mu_Q\longrightarrow \mu_Q^{-1}$
as compared to (\ref{evs}) and multiplied by an overall factor $A^{4(2k+1)}$
to provide the answer in topological framing \cite{univ1}: 
\be
{\cal H}_{\rm adjoint}^{{\rm Torus}_{[2,2p+1]}}(q,A) =
\frac{\{q\}^2}{\{Aq\}\{A/q\}} \left(
A^{8p+4}  + (qA)^{4p+2}\,\frac{\{Aq^3\}\{A\}^2\{A/q\}}{[2]^2\{q\}^4}
+ (q^{-1 }A)^{4p+2}\,\frac{\{Aq\}\{A\}^2\{A/q^3\}}{[2]^2\{q\}^4}
- \right.\nn\\ \left.
- 2\cdot A^{4p+2}
\frac{\{Aq^2\}\{Aq\}\{A/q\}\{A/q^2\}}{[2]^2\{q\}^4}
\right) \ \ \ \ \ \ 
\label{torusadjoint}
\ee
In this form, however, obscure is the differential-expansion structure \cite{diffexpan},
which is important to explain the cancelation of $\{A\}^2$ in denominator of (\ref{satfvsadj}).
One can rewrite this formula as
\be
\boxed{
{\cal H}_{\rm adjoint}^{{\rm Torus}_{[2,2p+1]}}(q,A) =1 + \{A\}^2\cdot\left(
\frac{q^2A^2\cdot \{A/q^3\}}{[2]^2\{q\}^2\cdot \{A/q\}}\cdot \Big(q^{4p}A^{4p} - 1\Big)
+ \frac{q^{-2}A^2\cdot \{Aq^3\}}{[2]^2\{q\}^2\cdot \{Aq\}}\cdot \Big(q^{-4p}A^{4p} -1\Big)
- \right.}\nn \\ \boxed{ \left.
- \frac{2A^2\cdot \{Aq^2\}\{A/q^2\}}{[2]^2\{q\}^2\cdot \{A\}^2}\cdot\Big(A^{4p}-1\Big) +
\frac{A^4\cdot\{q\}^2}{\{Aq\}\{A\}^2\{A/q\}}\cdot \Big(A^{8p}-1\Big)
\right)}
\label{torus2adjoint}
\ee
where the combination of four terms in braces is actually a Laurent polynomial,
but this is not transparent and therefore not very useful.

For generic twist knot one needs a new calculation, which gives
\be
\boxed{
{\cal H}_{\rm adjoint}^{{\rm Twist}_p}(q,A)
= 1 + \{A\}^2\cdot
\frac{A^4}{(q+q^{-1})^2}\cdot\left(-2(q+q^{-1})(q^3+q^{-3})\,\frac{A^{2p}-1}{A^2-1}
+ \right. }  \nn \\  \boxed{ \left.
+ {\{Aq^3\}(Aq^{-1}+qA^{-1})} \, \frac{q^{4p}A^{4p}-1}{q^2A^2-1}
+  {\{Aq^{-3}\}(Aq+A^{-1}q^{-1})} \,\frac{q^{-4p}A^{4p}-1}{q^{-2}A^2-1}
- 2\{Aq^2\}\{Aq^{-2}\}\,\frac{A^{4p}-1}{A^2-1}
\right)}
\label{twistadjoint}
 \\ \nn \\
= 1 + \{A\}^2\cdot
 {A^4} \cdot
\left(-2 (q^2-1+q^{-2})\sum_0^{p-1} A^{2s}
+ \right.   \nn \\   \left.
+\sum_{s=0}^{2p-1} A^{2s}\cdot
\frac{ A^2\,(q^{s+1}-q^{-s-1})^2 + q^4\,(q^{2s}+2-q^{-2s}) +
q^{-4}(-q^{2s}+2+q^{-2s}) - A^{-2}(q^{s-1}+q^{-s+1})^2}{(q+q^{-1})^2}
\right)
\nn
\ee
This is also a Laurent polynomial, i.e. the numerator is divisible
by $[2]^2=(q+q^{-1})^2$ after summation over $s$.
For example, the highest term with $s=2p-1$ in the numerator is just 
$(q^{2p}-q^{-2p})^2 = (q^p+q^{-p})^2\cdot (q^p-q^{-p})^2$ -- and if $p$ is odd,
then the first factor is divisible by $[2]$, otherwise, if $p$ is even,
the second factor is further factorized into $(q^{p/2}+q^{-p/2})^2\cdot (q^{p/2}+q^{-p/2})^2$,
and either $p/2$ is odd and $[2]$ divides the first factor, or the second factor
is further factorized, and so on. 

The {\it special} 
uniform {\it adjoint} polynomials at  $q=1$ are full {\it squares}
of the fundamental ones, as usual \cite{univ1}:
\be
\sigma_{\rm adjoint}^{{\rm Torus}_{[2,2p+1]}}( A)=
{\cal H}_{\rm adjoint}^{{\rm Torus}_{[2,2p+1]}}(q=1,A)
=  A^{4p}\cdot\left(pA^2-p-1) \right)^2 
= \left(\sigma_{_{\Box}}^{{\rm Torus}_{[2,2p+1]}}( A)\right)^2=
\nn \\
= 1 + \{A\}^2\cdot \left(p^2A^{4p+2} - \frac{A^{4p} -2p(A^2-1)A^{4p}-1}{\{A\}^2}\right) 
\ee
\vspace{-0.5cm}
\be
\sigma_{\rm adjoint}^{{\rm Twist}_p}( A)= 
{\cal H}_{\rm adjoint}^{{\rm Twist}_p}(q=1,A)
=  \left(A^{2p+2}-A^{2p}-A^2 \right)^2 
= \Big(\sigma_{_\Box}^{{\rm Twist}_p}( A)\Big)^2 
= \nn \\
= 1 + \{A\}^2\cdot \left(A^{2p+2}\cdot\left(A^{2p}+A^{-2p}\right)
+ 2\cdot \frac{1-A^{2p+2}}{1-A^{-2}}\right)
\ee

\section{HOMFLY for twisted satellites of 2-strand torus and twist knots}

Making use of these formulas we obtain for $\ {\cal K}=3_1={\rm trefoil}\ $:
\vspace{-0.2cm}
\be
H^{S_k(3_1)}_{\Box} =
\overbrace{1+\{Aq\}\{A/q\}\left(\frac{1-A^{2k}}{1-A^{-2}}\right.}^{H^{{\rm Twist}_{k }}_{_\Box}}
\left. \phantom{\frac{ A^{2 }}{ A^{2}}}\!\!\!\!\!\!\!\!\!
+ A^{2k} \Big((q^2+q^{-2})^2 - (q^4-q^2+1-q^{-2}+q^{-4})A^2-\{q\}^2A^4\Big)\right)
\label{Sk31f}
\ee
Note that $k$ here is actually $k+w^{3_1'}+3=k$,
for alternative expression with explicit $k+w$ see (\ref{twkSATto2p}) below.
Similarly, for $\ {\cal K}=4_1={\rm figure\ eight}\ $
with $w^{4_1'}=0$
\vspace{-0.2cm}
\be
H^{S_k(4_1)}_{\Box} =  
\overbrace{1+ \{Aq\}\{A/q\}\,\frac{1-A^{2k}}{1-A^{-2}}}^{H^{{\rm Twist}_k}_{_\Box}}
- A^{2k+1}\{Aq\}\{A\}\{A/q\}  
\Big((1+\{q\}^2)^2(A^2+A^{-2}) + (q^4+1+q^{-4})\{q\}^2\Big) 
\ee
To this one can add the particular (fundamental) case  of (\ref{satunknot}):
\be
H^{S_k({\rm unknot})}_{_\Box} = H^{{\rm Twist}_k}_{_\Box} =
1 + \{Aq\}\{A/q\}\cdot \frac{A}{\{A\}}(1-A^{2k})
\label{Htw1}
\ee

Eqs.(\ref{torus2adjoint}), (\ref{twistadjoint})
and (\ref{satfvsadj})
provide the fundamental HOMFLY for arbitrary $k$-twisted satellite
of arbitrary 2-strand torus knot:
\be
\boxed{H^{S_k({\rm Torus}_{[2,2p+1]})}_{_\Box} 
 = H^{{\rm Twist}_{k+w}}_{_\Box}
- \{Aq\}\{A\}\{A/q\}\cdot
\frac{A^{2k+2w+5}}{(q+q^{-1})^2}\cdot
\left(\frac{(q+q^{-1})^2\cdot\{q\}^2}{\{Aq\}\{A\}^2\{A/q\}}\cdot \Big(A^{8p}-1\Big)
+ \right.}\nn \\  \! \boxed{ \left.
+\frac{q^2A^{-2}\cdot \{A/q^3\}}{ \{q\}^2\cdot \{A/q\}}\!\cdot\! \Big(q^{4p}A^{4p} - 1\Big)
+ \frac{q^{-2}A^{-2}\!\cdot \{Aq^3\}}{ \{q\}^2\cdot \{Aq\}}\cdot\! \Big(q^{-4p}A^{4p} -1\Big)
- \frac{2A^{-2}\cdot \{Aq^2\}\{A/q^2\}}{ \{q\}^2\cdot \{A\}^2}\!\cdot\!\Big(A^{4p}-1\Big)
\right)} \  
\label{twkSATto2p}
\ee
and of arbitrary twist knot:
\be
\boxed{H^{S_k({\rm Twist}_p)}_{_\Box} 
= H^{{\rm Twist}_{k+w}}_{_\Box}
- \{Aq\}\{A\}\{A/q\}\cdot 
\frac{A^{2k+2w+5}}{(q+q^{-1})^2}\cdot\left(-2(q+q^{-1})(q^3+q^{-3})\,\frac{A^{2p}-1}{A^2-1}
+  \right. } \ \ \ \ \nn \\  \boxed{ \left. 
+ {\{Aq^3\}(Aq^{-1}+qA^{-1})} \, \frac{q^{4p}A^{4p}-1}{q^2A^2-1}
+  {\{Aq^{-3}\}(Aq+A^{-1}q^{-1})} \,\frac{q^{-4p}A^{4p}-1}{q^{-2}A^2-1}
- 2\{Aq^2\}\{Aq^{-2}\}\,\frac{A^{4p}-1}{A^2-1}
\right)} \ \ \ \ 
\label{twkSATtwp}
\ee
and the special polynomials for these satellites are   just
\vspace{-0.2cm}
\be
\sigma^{S_k({\rm Torus}_{[2,2p+1]})}_{_\Box}(A) =
H^{S_k({\rm Torus}_{[2,2p+1]})}_{_\Box}(q=1,A) =
1 +    { (A^2-1)}\cdot\left(1- 
A^{2k}\cdot\left(pA-\frac{p+1}{A} \right)^2\right)
\label{sigSatto2}
\ee
\vspace{-0.5cm}

\noindent
with $k = k+\underbrace{w^{{\rm Torus}_{[2,2p+1]}'}}_{-(2p+1)}+2p+1$ \  
and
\vspace{-0.3cm}
\be
\sigma^{S_k({\rm Twist}_p)}_{_\Box}(A) =
H^{S_k({\rm Twist}_p)}_{_\Box}(q=1,A) =
1 +    { (A^2-1)}\cdot\left(1-A^{2k}\cdot
\Big(1-A^{-2}-A^{-2p}\Big)^2\right)
\label{sigSattw}
\ee
\vspace{-0.2cm}

\noindent
with $k = k+\overbrace{w^{{\rm Twist}_{p}'}}^{-(2p+2)}+2p+2$. 
Note that because of the difference in write numbers in the 2-strand
torus and twist realisations we have for the 
trefoil = $3_1={\rm Torus}_{[2,3]}={\rm Twist}_1$
in the role of the companion 
\be
S_{k}({\rm Torus}_{[2,3]}) = S_{k+1}({\rm Twist}_{1})
\label{twotrefs}
\ee
with two different twist knots in the role of a pattern.

\section{A puzzle with superpolynomials
\label{supps}}

Superpolynomials \cite{GSV,DGR,DMMSS}
are the  {\bf t}-deformations of HOMFLY, depending on an extra parameter ${\bf t}=-q/t$,
which are positive Laurent polynomials
(i.e. all items enter with non-negative integer coefficients)
in the DGR variables \cite{DGR}
\vspace{-0.3cm}
\be
{\bf q}=t, \ \ \ \ \  {\bf t}=-q/t, \ \ \ \ \ {\bf a}^2 = A^2t/q \nn \\
A^2 = -{\bf a^2t}, \ \ \ \ q=-{\bf qt}, \ \ \ \ t={\bf q}
\label{dgr}
\ee
It could seem that eq.(\ref{SatH})  allows 
a straightforward deformation from HOMFLY to superpolynomials,
if positive adjoint superpolynomials are known for ${\cal K}$.

The celebrated example of such deformation is that of \cite{evo} for
twist-knots HOMFLY in symmetric and antisymmetric representations.
In the simplest case of the fundamental representation it is
\be
1+F(A^2,q^2)\{Aq\}\{A/q\} = 1+F(A^2,q^2))\Big(\{A\}^2 - \{q\}^2\Big) \ \longrightarrow\nn\\
\longrightarrow \
1 + F(A^2q/t,q^2)\left(\left\{A\sqrt{q/t}\right\}^2 - (q-t^{-1})(t-q^{-1})\right) =
1 + F(A^2q/t,q^2)\{Aq\}\{A/t\} = \nn \\
= 1+F({\bf a^2t^2},{\bf q^2t^2})\left\{{\bf a^2t^2} + {\bf q^2t}
+{\bf q^{-2}t^{-1}} {\bf a^{-2}t^{-2}}\right\}
\ee
In the case of twist knots $F(A^2,q^2) = -A^{k+1}\cdot \frac{\{A^k\}}{\{A\}}$.
For negative values of $k\leq 0$ this gives a positive polynomial, while for
$k>0$ we get instead a pure negative expression, with the first item $1$ canceled
with $-1$ coming from the second piece of the formula.
Note that for knots of other types $q^2$ can also require deformation:
the rule is not fully universal, see \cite{evo} for details.
Another remark is that the deformation
$\{q\}^2 \longrightarrow (q-t^{-1})(t-q^{-1})$ is exactly the one used in
\cite{AnoMMM21} to handle representation ${\it [21]}$ (which is actually the adjoint for $sl_3$),
and it is in conflict with the Hurwitz structure in the case of hyperpolynomials
\cite{MMSSsuppexpan}.

It can seem that eq.(\ref{satfvsadj}) 
in its second version is nicely consistent with the deformation to superpolynomials:
if $H-1$ and ${\cal H}-1$ are changed for positive superpolynomials 
for the pattern and companion,
the sum could also be interpreted as a positive superpolynomial for the satellite.
However, things are not so simple.
There are at least three immediate difficulties.

1) It is not quite so simple to find a positive polynomial deformation of  
(\ref{twistadjoint}), i.e. adjoint superpolynomials are a puzzle already for
the simplest possible twist knots in the role of companions.

2) The most naive deformation of the coefficient in (\ref{satfvsadj}) is  
{\it negative} rather than positive: 
\be
\frac{\{Aq\}\{A/q\}(1-A^2)}{\{A\}^2} \ \stackrel{(\ref{dgr})}{\longrightarrow} \ 
\frac{\{Aq\}\{A/t\}(1-A^2)}{\{A\}^2} = 
-\frac{({\bf a^2t^2}+{\bf q^2t} + {\bf q^{-2}t^{-1}}+{\bf a^{-2}t^{-2}})(1+{\bf a^2t})}
{{\bf a^2t} + 2 + {\bf a^{-2}t}}
\ee 

3) As we explained in the previous paragraphs the superpolynomials $H$ for the  twist knots, 
used as a pattern, are actually positive for $k\leq 0$ and get negative for $k>0$.
At the same time the last term in   (\ref{satfvsadj}) does not depend on $k$,
and can not change sign together with it.
As a simple  example we can take a naive ${\bf t}$-deformation of (\ref{Sk31f}):
\be
H^{S_k(3_1)}_{\Box} =
\underbrace{1+\{Aq\}\{A/q\}\left(\frac{1-A^{2k}}{1-A^{-2}}\right.}_{H^{{\rm Twist}_{k }}_{_\Box}}
\left. \phantom{\frac{ A^{2 }}{ A^{2}}}\!\!\!\!\!\!\!\!\!
+ A^{2k} \Big((q^2+q^{-2})^2 - (q^4-q^2+1-q^{-2}+q^{-4})A^2-\{q\}^2A^4\Big)\right)
\ \ \longrightarrow
\nn 
\ee 
\vspace{-1.2cm}
\be
\vspace{-0.6cm}
\longrightarrow \ \ \ \ \ \ \ \ \ \ \ \ 
P^{S_k(3_1)}_{\Box} \ \stackrel{?}{=} \
\overbrace{
1+\frac{({\bf a^2q^2t^3}+1)({\bf a^2t} + {\bf q^2})}{{\bf a^2q^2t^2}}
\left\{\ \boxed{\frac{1-({\bf a t})^{2k}}{1+({\bf at})^{-2}}}
\right.}^{P^{{\rm Twist}_{k }}_{_\Box}}\ + \ \ \ \ \ \ \ \ \ \
\nn \\
\left. \phantom{\boxed{\frac{ A^{2 }}{ A^{2}}}}\!\!\!\!\!\!\!\!\!
+ ({\bf a^2t})^k\cdot \Big(({\bf q^2t^?}+{\bf q^{-2}t^{-?}})^2 
+ ({\bf q^4t^2}+{\bf q^2t}+1+{\bf q^{-2}t^{-1}}+{\bf q^{-4}t^{-4}})\cdot{\bf a^2t}
+\frac{({\bf q^2t}+1)^2}{{\bf q^2t}}\cdot {\bf a^4t^2}\Big)\right\}
\label{Sk31Pf}
\ee
and the boxed ratio is a positive/negative polynomial depending on the sign of $k$.
This means that the r.h.s.  can actually contain terms of different signs,
while the l.h.s. is a polynomial for satellite in
the fundamental re\-presen\-tation,
and there is no doubt that pure positive (or pure negative) superpolynomials 
exists in this case.

Thus we are left with just a few options:

(a) either there is a delicate cancellation at the r.h.s.  
making the entire expression pure positive or negative,

(b) or uniform adjoint superpolynomials are not positive,
what my be not so surprising, because adjoint is a non-rectangular representation,

(c) or (\ref{satfvsadj})  does not survive the ${\bf t}$-deformation.

It remains for the future to understand what is actually true.
Whatever will be the outcome, there is a well defined problem to build
the fundamental superpolynomials for satellites -- if (\ref{satfvsadj})
does not help, one should look for alternative ways to solve it.


\section{Conclusion}

In this letter we originated the study of knot polynomials for twisted satellite knots.
This is a very interesting class, providing a new way to extend the results
for twist \cite{twistresults} and arborescent \cite{arbor} knots to far more complicated
knots.

An interesting part of the story is that the answers are naturally sums over representations
from $Q\in R\otimes\bar R$, which include adjoint representation and its descendants,
forming the so called $E_8$ sector of representation theory,
which is known \cite{univ1,univ2} to respect Vogels's universality \cite{Vogel},
i.e. can be described as a 3-parametric family, where all simple gauge algebras
appear at particular values of these parameters.
Especially amusing is that these universal expressions involve the somewhat
peculiar {\it uniform} polynomials \cite{univ1} (a kind of projective limit) --
still they enter formulas for ordinary HOMFLY of satellites, beginning from the
fundamental ones.
Since these latter ones are perfectly well defined,
we get a new way to test the relevance of uniform HOMFLY
-- however the simplest of the satellites has 14 crossings
and  additional work is needed to directly calculate its fundamental HOMFLY.
 
Another direction, open by our study, concerns superpolynomials.
This time it is important the adjoint representations of $sl_N$ algebras
are non-rectangular $[21^{N-2}]$ -- and it is a long-standing problem what
are the superpolynomials in non-rectangular case, do they exist at all and
are they {\it positive}.
At least the {\it hyperpolynomials} \cite{hyperpols}, which seem best respecting the
algebraic structures, are known to be non-positive beyond rectangular
representations, see \cite{AnoMMM21} and \cite{MMSSsuppexpan}
for detailed exposition of the problem.
If, despite the problems summarized in sec.\ref{supps}, 
decomposition like (\ref{satfvsadj})
of knot polynomials for satellites can be lifted to superpolynomials,
this would open an additional channel to study the problem --
at least for {\it uniform} superpolynomials from adjoint family.

In this letter we made just the first steps along this new path,
and there are clearly many more to do.
Most important, we now have all the tools for further investigation --
the next steps look tedious, but doable.
We hope that new results will follow in the near perspective.

\section*{Acknowledgements}

\noindent
I am indebted to Qingtao Chen for attracting my attention to twisted satellites.

\noindent
This work was funded by the Russian Science Foundation (Grant No.16-12-10344).


\begin{thebibliography}{12}



\bibitem{knotpols}
J.W.Alexander, Trans.Amer.Math.Soc. 30 (2) (1928) 275-306\\
V.F.R.Jones, Invent.Math. 72 (1983) 1 Bull.AMS 12 (1985) 103 Ann.Math. 126 (1987) 335\\
L.Kauffman, Topology 26 (1987) 395\\
P.Freyd, D.Yetter, J.Hoste, W.B.R.Lickorish, K.Millet, A.Ocneanu, Bull. AMS. 12 (1985) 239\\
J.H.Przytycki and K.P.Traczyk, Kobe J Math. 4 (1987) 115-139\\
A.Morozov, Theor.Math.Phys. 187 (2016) 447-454, arXiv:1509.04928

\bibitem{CS}
S.Chern and J.Simons, Proc.Nat.Acad.Sci. 68 (1971) 791–794;
Annals of Math. 99 (1974) 48-69; \\
A.S.Schwarz, New topological invariants arising in the theory of quantized fields,
Baku Topol. Conf., 1987\\
E.Witten, Comm.Math.Phys. 121 (1989) 351;\\
M.Atiyah,
{\it The geometry and physics of knots}, (CUP, 1990)\\
E.Guadagnini, M.Martellini and M.Mintchev, Clausthal 1989, Procs.307-317; 
 Phys.Lett. B235 (1990) 275\\
N.Reshetikhin and V.Turaev, Comm. Math. Phys. 127 (1990) 1-26

\bibitem{sat} 
M.Hedden, math/0606094

\bibitem{cabling}
A.Anokhina and An.Morozov,
Theor.Math.Phys. 178 (2014) 1-58, arXiv:1307.2216  

\bibitem{MMM2} A.Mironov, A.Morozov and An.Morozov,
 JHEP 03 (2012) 034,  arXiv:1112.2654



\bibitem{evo} A.Mironov, A.Morozov and An.Morozov,
AIP Conf. Proc. 1562 (2013) 123, arXiv:1306.3197

\bibitem{univ1}
A.Mironov, R. Mkrtchyan and A.Morozov,
Journal of High Energy Physics 02 (2016) 78,
 arXiv:1510.05884

\bibitem{DMMSS}
 P.Dunin-Barkowski, A.Mironov, A.Morozov, A.Sleptsov, A.Smirnov,
 JHEP 03 (2013) 021, arXiv:1106.4305

\bibitem{univ2}
A.Mironov and  A.Morozov,
Physics Letters B755 (2016) 47-57, arXiv:1511.09077

\bibitem{knotebook} http://knotebook.org

\bibitem{arbor}
A.Mironov, A.Morozov, An.Morozov, P.Ramadevi, V.K. Singh, JHEP 1507 (2015) 109, arXiv:1504.00371\\
S.Nawata, P.Ramadevi and Vivek Kumar Singh, arXiv:1504.00364\\
A.Mironov and A.Morozov, Phys.Lett. B755 (2016) 47-57, arXiv:1511.09077\\
A. Mironov, A. Morozov, An. Morozov, P. Ramadevi, V.K. Singh and A. Sleptsov,
J.Phys. A: Math.Theor. 50 (2017) 085201, arXiv:1601.04199 

\bibitem{RJ}
M.Rosso and V.F.R.Jones, J.Knot Theory Ramifications, 2 (1993) 97-112\\
X.-S.Lin and H.Zheng, Trans.Amer.Math.Soc. 362 (2010) 1-18 math/0601267\\
M.Tierz, Mod. Phys. Lett. A19 (2004) 1365-1378, hep-th/0212128\\
A.Brini, B.Eynard and M.Marino, Annales Henri Poincare, 13 (2012) No.8,  arXiv:1105.2012 \\
A.Alexandrov, A. Mironov, A.Morozov, An.Morozov, 
JETP Letters 100 (2014) 271-278, arXiv:1407.3754

\bibitem{diffexpan}
H.Itoyama, A.Mironov, A.Morozov and An.Morozov, JHEP 2012 (2012) 131, arXiv:1203.5978 \\
S.Arthamonov, A.Mironov, A.Morozov and An.Morozov, JHEP 04 (2014) 156, arXiv:1309.7984 \\
Ya.Kononov and A.Morozov, JETP Letters 101 (2015) 831-834, arXiv:1504.07146 \\
Q.Chen, arXiv:1512.07906 \\
C.Bai, J.Jiang, J.Liang, A.Mironov, A.Morozov, An.Morozov and A.Sleptsov, arXiv:1709.09228

\bibitem{GSV}
S.Gukov, A.Schwarz and C.Vafa, Lett.Math.Phys. 74 (2005) 53-74, arXiv:hep-th/0412243

\bibitem{DGR} N.M.Dunfield, S.Gukov, and J.Rasmussen, math/0505662


\bibitem{AnoMMM21}
A.Anokhina, A.Mironov, A.Morozov and An.Morozov,
 Nucl.Phys.B 882C (2014) 171-194,  arXiv:1211.6375
 
\bibitem{MMSSsuppexpan} 
A.Mironov, A.Morozov, A.Sleptsov and A.Smirnov,
     Nucl.Phys. B 889 (2014) 757-777, arXiv:1310.7622

\bibitem{twistresults}
A.Morozov, Nucl.Phys. B911 (2016) 582-605, arXiv:1605.09728;
  JHEP 1609 (2016) 135, arXiv:1606.06015;
  arXiv:1612.00422; Phys.Lett. B 766 (2017) 291-300, arXiv:1701.00359; 
  arXiv:1711.09277;  Mod.Phys.Lett. A  (2018), arXiv:1712.03647\\ 
Ya.Kononov and A.Morozov, Theor.Math.Phys. 193 (2017) 1630-1646, arXiv:1609.00143;
  Mod.Phys.Lett. A31 (38) (2016) 1650223, arXiv:1610.04778
 



\bibitem{Vogel} B. Kostant, Proc. Natl. Acad. Sci. USA, Mathematics 81 (1984) 5275-5277\\
P.Deligne, C.R.Acad.Sci. 322 (1996) 321-326 \\
P.Deligne, R.de Man, C.R.Acad.Sci. 323 (1996) 577-582 \\
A. Cohen, R.de Man, C.R.Acad.Sci. 322 (1996) 427-432 \\
P.Vogel, {\it The universal Lie algebra} (1999),
http://webusers.imj-prg.fr/pierre.vogel/\,;
{\it Algebraic structures on modules of diagrams} (1995);
J.Pure Appl. Algebra 215 (2011) 1292-1339 \\
J.Landsberg, L.Manivel, math.AG/0203241; Adv.Math. 171 (2002) 59-85;
  Adv. Math. 201 (2006) 379-407 \\
R.Mkrtchyan and A.Veselov, JHEP 1208 (2012) 153, arXiv:1203.0766 \\
R.Mkrtchyan, Phys.Lett. B105 (1981) 174 \\
P.Cvitanovic, {\it Group Theory}, Princeton University Press, 2004,
http://www.nbi.dk/grouptheory

\bibitem{hyperpols}
M.Aganagic and Sh.Shakirov, 
Commun. Math.Phys. 333(1)  (2015) 187-228, arXiv:1105.5117 \\
I.Cherednik, arXiv:1111.6195 \\
E.Gorsky, S.Gukov and M.Stosic, arXiv:1304.3481       \\
S.Arthamonov, A.Mironov, A.Morozov,
Theor.Math.Phys. 179 (2014) 509-542,  arXiv:1306.5682  \\
I.Cherednik and I.Danilenko, arXiv:1408.4348 \\
S.Arthamonov and Sh.Shakirov,  arXiv:1504.02620;  arXiv:1704.02947 \\
S.Nawata and A.Oblomkov,  arXiv:1510.01795   


\end{thebibliography}
\end{document}